\documentclass[10pt,letterpaper]{article}
\usepackage[top=0.85in,left=2.75in,footskip=0.75in]{geometry}

\usepackage{amsmath,amssymb}

\usepackage{changepage}

\usepackage[utf8]{inputenc}

\usepackage{textcomp,marvosym}

\usepackage{cite}

\usepackage{nameref,hyperref}

\usepackage[right]{lineno}

\usepackage{microtype}
\DisableLigatures[f]{encoding = *, family = * }

\usepackage[table]{xcolor}
\usepackage{xspace}
\usepackage{subcaption}
\usepackage{multirow,multicol,colortbl,tabularx}

\usepackage{array}

\newcolumntype{+}{!{\vrule width 2pt}}

\newlength\savedwidth

\raggedright
\usepackage[aboveskip=1pt,labelfont=bf,labelsep=period,justification=raggedright,singlelinecheck=off]{caption}

\makeatletter
\renewcommand{\@biblabel}[1]{\quad#1.}
\makeatother

\usepackage{lastpage,fancyhdr,graphicx}
\usepackage{epstopdf}
\fancyheadoffset[L]{2.25in}
\fancyfootoffset[L]{2.25in}
\begin{document}
\vspace*{0.2in}

\begin{flushleft}
{\Large
\textbf\newline{Touchtone leakage attacks via smartphone sensors: mitigation without hardware modification}

}

Connor Bolton\textsuperscript{1},
Yan Long\textsuperscript{1},
Jun Han\textsuperscript{2},
Josiah Hester\textsuperscript{3},
Kevin Fu\textsuperscript{1}

\textbf{1} College of Computer Science and Engineering, University of Michigan, Ann Arbor, Michigan, USA
\\
\textbf{2} College of Computer Science, National University of Singapore, Singapore
\\
\textbf{3} College of Computer Science, Northwestern University, Evanston, Illinois, USA
\\


\end{flushleft}
\section*{Abstract}
Smartphone motion sensors provide a concealed mechanism for eavesdropping on acoustic information, like touchtones, emitted by a device.
Eavesdropping on touchtones exposes credit card information, banking pins, and social security card numbers to malicious 3rd party apps requiring only motion sensor data.
This paper's primary contribution is an analysis rooted in physics and signal processing theory of several eavesdropping mitigations, which could be implemented in a smartphone update. We verify our analysis imperially to show how previously suggested mitigations, i.e. a low-pass filter, can undesirably reduce the motion sensor data to \textit{all applications} by 83\% but only reduce an advanced adversary's accuracy by less than one percent. Other designs, i.e. anti-aliasing filters, can fully preserve the motion sensor data to support benign application functionality while reducing attack accuracy by 50.1\%. 
We intend for this analysis to motivate the need for deployable mitigations against acoustic leakage on smartphone motion sensors, including but not limited to touchtones, while also providing a basis for future mitigations to improve upon.




\section{Introduction}

\begin{figure}[!h]
    \centering
    \includegraphics[width=.95\textwidth]{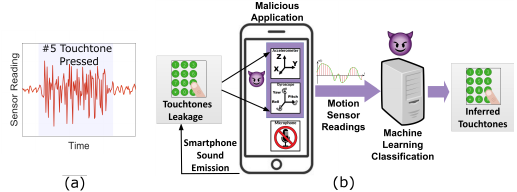}
	\caption{\textbf{Touchtone leakage and eavesdropping.} (a) A touchtone, indicating a ``5" on a smartphone number pad, leaks into accelerometer data. (b) A malicious smartphone application can classify this leakage to descern that a ``5" touchtone was emitted, infering user input of a ``5" for purposes such as dialing a phone number or inputing information into automated services.}
    \label{fig:intro_figure}
\end{figure}

Touchtones, the sounds produced by a smartphone when a numerical key is pressed, are an established communication standard~\cite{dtmfStandard} often used to encode user feedback in telephony channels.
In modern telephony systems, touchtones often represent important information such as credit card numbers (during activation), bank pins, various account numbers, social security numbers, selections for various options in an automated services, and possibly even votes in a federal election (done by phone)~\cite{vasilogambros2019voting}.
Recent research has shown that sound produced by a smartphone's speaker may ``leak" into the same phone's motion sensors, particularly speech.
Our experiments show that \textit{touchtone leakage}, touchtone information leaking into motion sensor data, occurs with a signal-to-noise ratio sufficient to be seen visibly~(Fig~\ref{fig:intro_figure}b).
This leakage enables malicious smartphone applications with motion sensor access (e.g., a seemingly benign smartphone game) to ascertain any numerical user input that produces touchtones (Fig~\ref{fig:intro_figure}), an attack we term \textit{touchtone eavesdropping}.
However, reducing acoustic leakage remains an open research problem as papers investigating this topic~\cite{gyrophone,accelword,anand2018speechless,anand2019spearphone} focus more on adversarial and threat modeling than mitigation efforts.
This paper's primary goal is to open discussion on how to reduce acoustic leakage into nearby motion sensors, using touchtone leakage as an exemplary case study.

To understand how to mitigate touchtone eavesdropping, we assess why acoustic information is hidden in motion sensor data and how signal processing and physical phenomenon, such as aliasing or varying frequency responses, aid adversarial recovery of the original user key press.
These phenomenon cause artifacts of touchtone information to manifest in a multitude of ways, but an adversary only needs to be able to ascertain user input through one of those manifestations.
More advanced techniques such as selective integration of multiple sensors and sensor axes via machine learning can instead utilize several of these manifestations simultaneously for a more proficient attack. Our experiments show user input can be recovered by an adversary at over 99~\% accuracy with such methods. 

This paper uses the above assessment to inform and analyze functionality-aware, software-updatable mitigation designs for touchtone leakage. 
Mitigations can reduce touchtone leakage by reducing the total information in sensor output, but this also affects benign applications relying on such data. It is thus important to keep functionality in mind when designing mitigations. Additionally, we focus on solutions that may be implemented as a software update to support existing devices and designs where hardware changes may not be viable.
Using these criteria we analyze both ineffective and effective solutions to demonstrate ideas to emulate or avoid. 
For example, we analyze and evaluate how some apparent mitigations briefly suggested in related work, such as sampling rate reduction or digital low-pass filtering alone, are ineffective at reducing touchtone leakage; sampling rate reduction can reduce available information to \textit{all} applications by more than 80\% yet our classifier maintains accuracy over 95\% for three of the four tested phones. 
Other designs, such as a software anti-aliasing filter that uses oversampling, do not change the amount of information available to applications while reducing accuracy by over 50.1\%.
Our contributions include:

\subsection{Contributions}
\noindent \textbf{Touchtone eavesdropping assessment:}
We discuss and experimentally demonstrate the relevant physics and signal processing theory of touchtone leakage to reveal challenges mitigations must consider, such as aliasing and non-linear frequency responses. We detail both simple and advanced adversaries, discussing how selective integration of multiple sensors' data with machine learning can further impede mitigation efforts.

\noindent \textbf{Defense design analysis:}
We analyze the advantages and disadvantages of several signal-processing based leakage reduction solutions. We explicitly include the idea of not reducing functionality as a design criteria. Some apparent approaches, such as reduced sampling rates, are nonintuitively ineffective mitigations. Other filter designs, such as a software Butterworth anti-aliasing filter, can reduce leakage in the signals as a software update. Hardware changes could serve as long term solutions.

\noindent \textbf{Implementation and evaluation:}
We implement and evaluate both simple and more advanced touchtone leakage attacks to serve as baselines for mitigation evaluation, then the mitigations themselves.
Baseline attacks can achieve accuracy higher than 99\%. We evaluate several signal processing mitigation designs to demonstrate both effective and ineffective designs. Apparent mitigations, such as digital low-pass filters and reduced sampling rates, may remain ineffective (less than a 1\% difference in accuracy from baseline) even while reducing benign information that may be crucial for applications to function. Our anti-aliasing filter reduces accuracy by over 50.1\% with no information loss, and can reduce further with minor benign information loss.
\section{Background and related work} \label{sec:background}

\subsection{Touchtones}
\begin{figure}[!h]
    \centering
    \includegraphics[width=.6\textwidth]{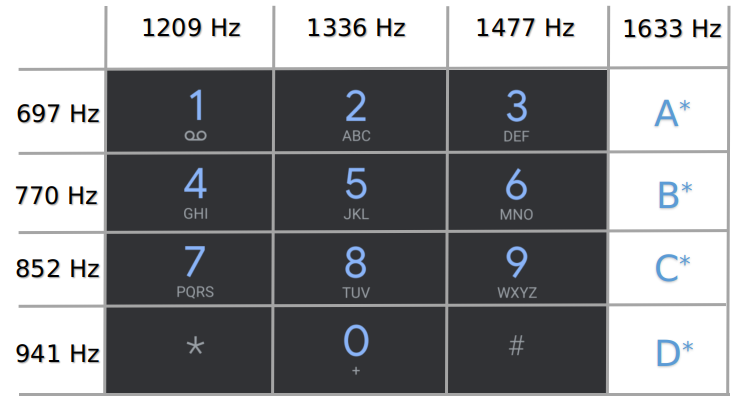}
    \caption{\textbf{Touchtone frequencies.} Touchtones are comprised of two single-frequency tones emitted simultaneously to convey numerical input.}
    \label{fig:dialtone-frequencies}

\end{figure}

Touchtones, also known as dual-tone multi-frequency (DTMF) signals, are a standardized~\cite{dtmfStandard} code of two-tone audible acoustic signals that play upon a numerical key press, often used in telecommunications or other applications with a numerical touch pad \cite{cho2008remote, ladwa2009control, sharma2006dtmf, grover2009hiv}. The sound produced by a phone when you press an individual key to dial a phone number, answer an automated telephony question (e.g. ``press 1 to...."), register credit card numbers or bank pins over the phone, or other such actions are examples of touchtones. There are 16 unique touchtones~(Fig~\ref{fig:dialtone-frequencies}), each consisting of two frequencies taken from two separate frequency sets, used for the numbers 0-9, the symbols * and \#, and an additional 4 tones reserved for special services. As they are unique, hearing one touchtone is indicative of a certain number press. These dual-tone combinations have been chosen specifically to be easily understood in the presence of noise for reliable communication.

\subsection{Signal Processing} \label{sec:background-signal-processing}
\subsubsection{Aliasing}
Aliasing can have several definitions depending on context, but the most relevant definition in the context of this paper refers to distortions caused by improper sampling of a signal \cite{maciejewski2009nonuniform, tsividis2004digital}.
As defined by the Nyquist sampling theorem~\cite{nyquist_rate,shannon} the highest frequency a sensor with sampling rate $f_s$ can properly sample is the Nyquist frequency $f_N=f_s/2$. If a signal has frequencies greater than $f_N$ the sensor output \textit{will} contain aliases of the original signal. The formula for the frequency of the alias, $f_a$, given the Nyquist frequency $f_N$ and the frequency of the original signal $f$ is $f_a=|2mf_N-f|$.

\subsubsection{Bandwidth}
The signal path of physical signals can normally only allow signals in a certain range of frequency to pass without strong attenuation because of the signal path's frequency response which is determined by its physical properties. Generally, the difference between the upper and lower bounds of such frequency range is defined as the bandwidth of the signal path. For an ideal motion sensor (with a flat frequency response) with the sampling rate of $f_s$, the its bandwidth is the same as the Nyquist frequency $f_N$ since only motion signals under the Nyquist frequency can be properly sampled and passed to the processor of smartphones. Although the nominal bandwidth is pre-determined by the properties of the signal path, people usually reduce the actual bandwidth by means of filtering. 

\subsubsection{Filtering}
Filtering is the process of reducing the bandwidth of a signal path by blocking (attenuating) unwanted signals at certain frequencies and only allowing desired signal to reach the destination, i.e., the smartphone processor in the context of this paper \cite{winder2002analog}. Common filters include low-pass filters, which blocks high-frequency signals and passing through high-frequency signals, and high-pass filters which achieve the opposite. The band of frequencies that the filters let pass through is called the pass band. Filters can be implemented either in software or in hardware, and can be implemented in different forms such as simple RC impulse response filters, Butterworth filters, Chebyshev filters, etc. Different implementations of filters have different frequency responses, meaning the abilities of blocking and passing signals at different frequencies are different. Ideally, the frequency response in the filters' pass band should be flat, so that the desired signals won't be distorted. However, such distortions are usually unavoidable for both software and hardware filters in the real world.

\subsection{Related work}
\textbf{Acoustic eavesdropping using motion sensors.} Previous works have shown the feasibility of acoustic eavesdropping attacks using motion sensor similar to touchtone eavesdropping in this paper. Gyrophone~\cite{gyrophone} demonstrates an attack that recognizes human-spoken digits using smartphone gyroscopes by extracting speech spectral information. Similarly,  Spearphone~\cite{anand2019spearphone} and AccelEve \cite{ba2020learning} uses smartphone accelerometers to eavesdrop human-spoken digits. AccelWord~\cite{accelword} investigate the feasibility of leveraging smartphone’s accelerometer to capture acoustic signals for low-power hotword detection. PitchIn~\cite{pitchin} fuses across multiple uni-model sensors (e.g., only accelerometers or gyroscopes) to reconstruct intelligible human speech by interleaving sensor readings from multiple sensors to increase the effective sampling rate. This work differ from the previous works in that: 1) We assess the privacy issue of eavesdropping touchtone information from smartphones, which require a different analysis methodology than the previous acoustic eavesdropping targets. 2) We analyze and evaluate the effectiveness as well as feasibility of several mitigations that can be practically implemented, and open up the discussion of future functionality-aware mitigations. 3) We inspect how acoustic leakage can manifest itself differently in separate axes' sensor readings of even a single sensor and uncover the fact that an advanced attacker may combine multidimensional information from different axes to enhance the attack. 

\textbf{Other motion sensor side-channels.}  Recent research also demonstrates novel side-channel attacks utilizing smart-phone motion sensors to infer victims’ location or keystrokes. ACComplice~\cite{accomplice} leverages smart-phone accelerometer to infer victim driver’s driving routes as well as starting point. Narain et al. further extended findings of ACComplice and demonstrated the feasibility of such attacks in large scale across ten cities~\cite{narain2016sp}. (sp)Iphone~\cite{spiphone} accesses acceleromter readings to infer typed text on nearby keyboards by observing the relative physical position and distance between the smartphone and keyboards and the vibration detected. Similarly, ACCessory~\cite{accessory} utilizes an accelerometer to infer keystrokes as the victim user types on his/her smart-phone. Due to minute differences in taps, it is able to sufficiently infer the typed keys. Tapprints~\cite{tapprints} further extends the findings of ACCessory by incorporating both accelerometer and gyroscopes as well as conducting larger experiments at scale with more practical use case scenarios. These motion sensor side-channel attacks against locations and keyboard inputs complement our discussions on eavesdropping touchtone information in this paper.

\textbf{Acoustic injection attacks.} Previous work has also explored motion sensors’ susceptibility to vibrations caused by acoustic signals, namely to affect motion sensor readings via acoustic injection. For instance, Son et al. proposes an attack to bring down and crash drones only by acoustic injection as MEMS gyroscopes are vulnerable to acoustic noises at their resonant frequencies~\cite{son2015rocking}. Similarly, \cite{trippel2017walnut, tu2018injected} propose acoustic injection attack on MEMS accelerometers to manipulate the output of the sensors by injecting certain acoustic signals at their resonant frequencies. Unlike these attacks that inject acoustic signals to motion sensors, we demonstrate the feasibility of capturing privacy sensitive information such as touchtone from acoustic signals naturally emitted from victim smartphones.
\section{Touchtone eavesdropping assessment}
\begin{figure}[!h]
	\centering
        \includegraphics[width=.9\textwidth]{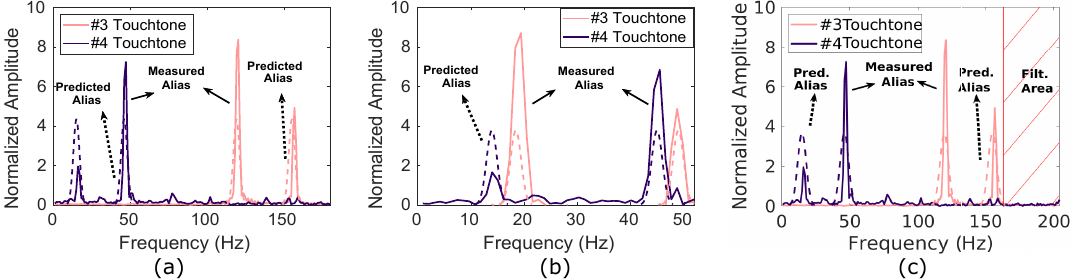}
	\caption{\textbf{Predictable and discernible touchtone leakage}. Touchtone leakage for \#3 and \#4 touchtones in a Google~Pixel~2's accelerometer's x-axis. These signals remain discernable and predictable in the frequency domain with (a) a normal, unaltered signal, and also despite previously suggested mitigations in (b) reduced sampling rates and (c) digital low-pass filtering.}
	\label{fig:dialtone-alias}
\end{figure}

For mitigation design it is crucial to understand how attacks utilizing touchtone leakage occurs~(Fig~\ref{fig:intro_figure}) and the multitude of reasons why it can be difficult to mitigate~(Fig~\ref{fig:dialtone-alias}). We assess (1) how touchtones produced by a phone's speaker leak distinguishable, deterministic byproducts to the smartphone's motion sensors (e.g. the accelerometer and gyroscope) and (2) how adversaries can use leakage to determine user input.

\subsection{Threat Model}
This paper considers an adversary whose goal is to determine a user's numerical key presses on a smartphone using access to a smartphone's motion sensor data and the knowledge of touchtone leakage, an attack we term \textit{touchtone eavesdropping}.
We assume the adversary can obtain and save motion sensor data through means such as a malicious application with motion sensor access. The adversary may have access to the same model as the victim's phone(s); a phone's model can be determined by an application using fingerprinting techniques~\cite{bojinov2014mobile,potharaju2012plagiarizing,zhang2012fingerprint}. The adversary can use their duplicate phone(s) to collect training data to build a classification system.
Last, the adversary has unlimited time to classify victim data as the victim data can be saved and sensitive information (e.g. credit card numbers, bank pins, social security numbers) may not change often.

\subsection{Touchtone information in motion sensor data}\label{sec:attack-causalities}
Touchtone leakage manifests itself in motion sensor data in a multitude of forms due to various physical and signal processing phenomenon. Each of these manifestations can contain  redundant or complementary information regarding the original touchtone. However, an attacker only needs one method, and possibly one manifestation of touchtone information, to achieve an eavesdropping attack. Defenders however, must consider how to block as much of this information as possible.

\subsubsection{Acoustic waves and sensor construction}
Acoustic waves produced by the smartphone's speaker alter the output of microelectricalmechanical systems (MEMS) accelerometers and gyroscopes \cite{beeby2004mems} due to how these sensors \textit{approximate} motion. MEMS accelerometers and gyroscopes approximate the motion of a larger body (i.e. a smartphone) via the motion of a small sensing mass(es) attached to capacitive springs. When the mass(es) moves, the springs create a representative voltage which is then amplified, filtered, digitized, and sent to the processor. However, while the linear or angular acceleration of the sensing mass(es) are usually accurate representations of the body's acceleration, they are not exact. For example, small acoustic vibrations via the air or contacted surfaces can move the small sensing masses even if minimally affecting the connected body (i.e. smartphone) due to effects such as varying frequency responses~\cite{gyrophone,trippel2017walnut,anand2019spearphone}. In this case, MEMS accelerometers and gyroscopes may capture acoustic signals.

\subsubsection{Touchtone aliasing} \label{sec:attack-aliasing}
Aliasing, described in Section~\ref{sec:background-signal-processing}, is a key factor in both making touchtone leakage occur and for making it difficult to mitigate.
Touchtones have frequencies higher than the Nyquist sampling rate for most smartphone motion sensors, and thus have aliases. However, the frequencies of these aliases can be predicted as the touchtone frequency and sampling rate are both known~(Fig~\ref{fig:dialtone-alias}). An attacker can use these known aliases to indicate the presence of the missing original touchtone frequencies.

Furthermore, the non-linear placement of these aliases --- how all touchtone frequencies can lie somewhere in the sampled signal's frequency band --- can make touchtone eavesdropping resistant to suggested mitigations. For example, reducing the sampling rate will not get rid of aliases, only move them~(Fig~\ref{fig:dialtone-alias}b). Low-pass filters may remain ineffective unless the cutoff frequency is placed low, as touchtone aliases could be close to 0~Hz (Fig~\ref{fig:dialtone-alias}c).

\subsubsection{A cacophony of sensitive information}~\label{sec:attack-information}
The two above factors enable touchtone leakage, but information leakage may manifest in a multitude of forms simultaneously (i.e. different axes of a sensor having different signals related to touchtones) due to a variety of physical and signal processing phenomenon; these manifestations can provide complementary or distinct information for the purpose of classifying touchtones (and thereby user input) and an attacker may need only one of these manifestations. 

\begin{figure}[!h]
    \centering
        \includegraphics[width=.95\textwidth]{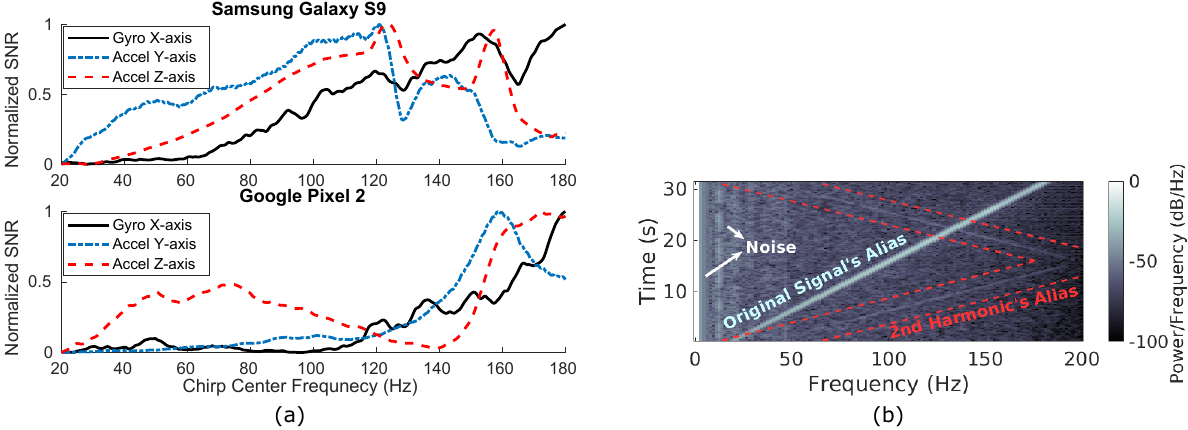}
    \caption{\textbf{Touchtone information manifestations.} Touchtone information can be embedded in a variety of forms or to varying extents in motion sensor data. In (a), two axes have distinct non-linear frequency responses to a 420~Hz to 580~Hz chirp from the loudspeaker. Different axes may be better predictors for certain tones. (b) shows how there may be many subtle artifacts in touchtone data. An attacker could use any of these artifacts to launch a touchtone eavesdropping attack.}
    \label{fig:snr_harmonics}
\end{figure}

Different sensors or sensor axes can contain complementary or different information about the same set of touchtones. One factor that can vary how information manifests is varying frequency responses in phone construction, speakers, sensors, or even individual axes of sensors. 
Different frequency responses inherent to physical materials and sensors can lead to one sensor axis having a higher signal-to-noise ratio (SNR) for certain frequencies (i.e. touchtones) where a separate axis could have a higher SNR for other frequencies~\cite{bolton2018blue,kune2013ghost,son2015rocking}~(Fig~\ref{fig:snr_harmonics}a). With access to both axes an adversary may be able to exploit this fact and combine useful information.

Additionally, in the same signal (i.e. sensor axis) information about the same touchtone can manifest in different manners. For example, an axis will have information on an alias of the touchtone frequency, but could also have information on the harmonics of the same touchtone~(Fig~\ref{fig:snr_harmonics}b). A touchtone eavesdropping attack would only need to recognize one of a touchtone's alias, harmonic, or even an alias of the harmonic to be successful.

\subsection{Adversarial touchtone recovery} \label{sec:attack-recovering}

The goal of the attacker is to recognize when a touchtone is pressed by discerning the presence of the eight individual touchtone frequencies~(Fig~\ref{fig:dialtone-frequencies}) in motion sensor data. The adversary can benefit by trying to use all possible touchtone information in motion sensor data, as discussed in Section~\ref{sec:attack-information}. The most straightforward approach to do this is by making a machine learning based classifier as the attacker does not particularly care which information the classifier uses, just that it can classify motion sensor data into touchtones. The advent of easily usable machine learning tools makes this task not arduous in the modern day.

Furthermore, adversaries can make use of the varying information in different sensors and sensor axes by selectively integrating data from multiple sensors (in our case the accelerometer and gyroscope). For example, one sensor axis may be more apt at discerning the presence of a particular touchtone but a separate sensor axis could be a better indicator of a separate touchtone. This same idea, using multiple sensors to reveal emergent information, has been used by researchers for benign purposes in several fields including on drones~\cite{lee2016drone}, body-sensor networks~\cite{gravina2017multi}, and much more. Building a classification model to specifically use this fact should  lead to more efficient attacks.

\section{Functionality-aware software mitigation design}
\label{sec:mitigation}

Touchtone eavesdropping mitigations require careful forethought and consideration of leakage mechanics~(Section~\ref{sec:attack-causalities}) to effectively reduce leakage while not 
hampering benign application behaviour. To accomplish this task, mitigations should reduce touchtone information in motion sensor data while minimally altering or reducing any other information. This paper adds an additional criteria that mitigations should be able to deploy as a software update to support current devices.
This section examines several mitigation designs to show how some apparent designs such as sampling rate reduction can only reduce touchtone leakage by reducing the total information in a signal, hampering application functionality, while other designs such as anti-aliasing filters can reduce touchtone leakage while minimally harming application functionality.

\subsection{Designing for both privacy and functionality} \label{sec:mitigation-goals}

While protecting the privacy of smartphone users from touchtone eavesdropping attacks is an urgent issue this paper is addressing, we consider ensuring functionality to be a second --- but no less critical --- criteria for mitigation design. The reason is that to be adopted into mainstream systems the mitigation must also support the expected functionality of motion sensor dependant applications. It is widely accepted that security and privacy must support some level of functionality and usability \cite{yan2019catcher, whitten1999johnny, matyavs2002biometric} as these features drive device markets and development.

Touchtone eavesdropping attackers and benign smartphone applications using the motion sensors use the same signals --- the motion sensor readings. As a result, limiting the attackers' capability might also inadvertently limit benign applications' performance. From a practical use standpoint, designing such mitigation for privacy protection thus also requires the designers to be functionality-aware and guarantee minimal degradation of functionality by carefully optimizing the implementation of their mitigation.

As discussed in Section~\ref{sec:background-signal-processing} two significant factors for reducing information in a signal --- and thereby reducing application functionality --- are (1) bandwidth reduction and (2) signal distortion; conversely, minimizing bandwidth reduction and signal distortion can better support application functionality. There is a near-unanimous trend of higher sensor bandwidth leading to higher performance in previous research in various activities such as  human activity recognition~\cite{khan2013exploratory}, animal health monitoring, \cite{pfau2021low}, road quality assessment \cite{bridgelall2015inertial}, etc. In addition, more commercialized techniques such as using motion sensor readings for smartphone image stabilization \cite{hu2016image} and rolling shutter correction \cite{karpenko2011digital} rely on sample rates higher than 100~Hz, correlating to a bandwidth of 50~HZ. Thus reducing bandwidth below those ranges risks causing these applications to malfunction and may stifle future application performance.
A significantly distorted signal could also impact application behavior and thus should also be minimized when possible, but it is more difficult to ascertain how much distortion is permissible. An ideal mitigation should support the original bandwidth with minimal distortion, only removing traces of touchtone byproducts.

\subsection{Apparent mitigations that sacrifice functionality}\label{sec:mitigation-bad}
Mitigation strategies predicated on reducing available sensor bandwidth may not only hinder application functionality, but also may ineffectively attenuate sensitive touchtone information. This section analyzes apparent mitigations of sampling rate reduction and digital low-pass filtering to show how touchtone information may persist despite significant bandwidth reduction.

\subsubsection{Sampling rate reduction}\label{sec:mitigation-reduced}
Lowering sampling rates directly lowers available bandwidth~(Section~\ref{sec:background-signal-processing}) in an attempt to also lessen the threat of acoustic eavesdropping; however, it is ineffective at attenuating leakage (Fig~\ref{fig:dialtone-alias}b) primarily due to how aliasing places touchtone information in a digital signal no matter the sampling rate (Section~\ref{sec:attack-aliasing}).
Even at very low frequencies, the eight touchtone frequencies still have aliases and thus leave discernible traces. Although with such an extremely low sampling rate, it might be difficult for an attacker to realistically differentiate between different touchtone aliases. But this also affects dependant application's functionality similarly. Our experiments back this intuition~(Section~\ref{sec:results-reduced}), as reduced sampling rates achieve minimal accuracy reduction for our touchtone eavesdropping attack until having sampling rates under 100~Hz, a forth of the original sampling rate.

\subsubsection{Digital Low-pass filter}\label{sec:mitigation-lpf}
Low pass filters may at first seem like a natural mitigation for touchtone leakage, which relies on aliasing, but a software digital low-pass filter alone cannot increase privacy while preserving functionality~(Fig~\ref{fig:dialtone-alias}c). To note, previous papers often do not specify which low-pass filter design they suggest, and hardware changes to include analog low pass filters may be a sufficient future defense as later discussed. However, when discussing software-updatable mitigations, digital low-pass filters alone also do not address the problem of aliasing. Referring back to Section~\ref{sec:attack-aliasing}, many of the resulting touchtone aliases could be under the low-pass filter cutoff frequency that is chosen due to the non-linear placement of alias frequencies. A lower cutoff frequency is more likely to attenuate more aliases, but only because it is reducing the available bandwidth for all motion sensor data. Thus it also suffers from needing to reduce available bandwidth to provide better privacy. Our experiments demonstrate this pathology~(Section~\ref{sec:results-lpf}).

\subsection{Designing functionality-aware signal processing mitigations} \label{sec:mitigation-good}

A functionality-aware signal processing mitigation should minimally reduce available bandwidth and distortion while attenuating touchtone leakage. Our approach is to rely on established digital signal processing techniques designed to eliminate specific leakage contributors, particularly aliasing.
We propose a software update enabling oversampling and digital anti-aliasing filters as a primary means of defense acoustic general acoustic leakage.
Additionally, we describe how one can utilize the predictable nature of touchtone aliases in defense design.

\subsubsection{Oversampling and digital anti-aliasing filters}\label{sec:mitigation-aa}
\begin{figure}[!h]
	\centering

        \includegraphics[width=\textwidth]{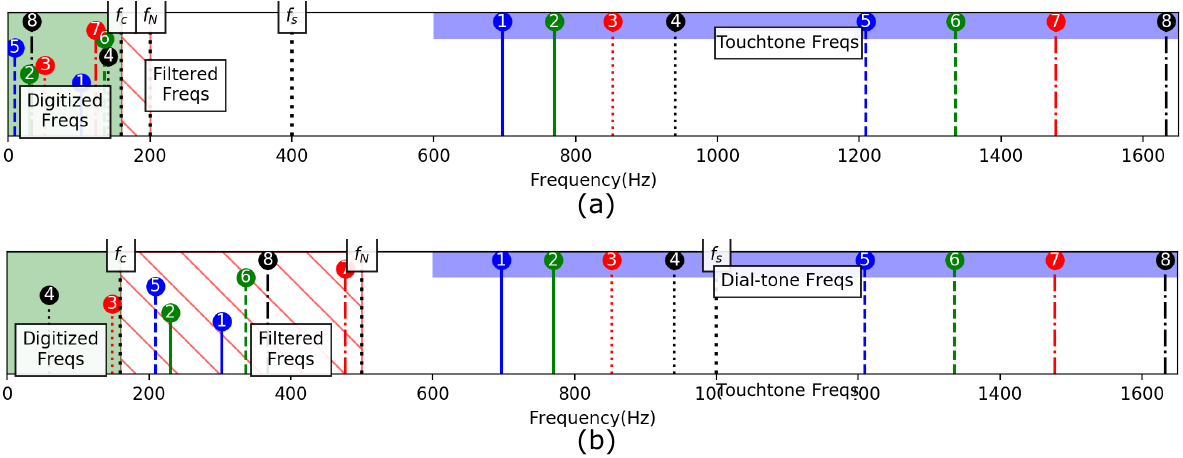}
	\caption{\textbf{A need for oversampling.} Digital anti-aliasing filters can attenuate more touchtone aliases than low-pass filter without reducing available bandwidth due to use of oversampling. In the example with sampling rate of $f_S=400Hz$, $f_N=200~Hz$, and $f_c=180~Hz$, touchtone aliases (numbered 1 to 8 to correlate with the eight touchtone frequencies in the blue box) are attenuated if filtered (diagonal red-lined area) and otherwise unattenuated (green area). (a) A digital low-pass filter may be unable to attenuate many touchtone frequency aliases without also eliminating significant frequency information benign applications may rely on (Section~\ref{sec:mitigation-lpf}). (b) A digital anti-aliasing filter with the same $f_c$ can filter more frequencies due to the use of oversampling (Section~\ref{sec:mitigation-aa}).}
	\label{fig:defense-freq-aa}
\end{figure}

Oversampling is the act of sampling at a faster rate than the bandwidth that you wish to eventually provide, and can be used to create anti-aliasing filters that reduce touchtone leakage while still providing the original bandwidth to current applications~(Fig~\ref{fig:defense-freq-aa}). Oversampling can be implemented as a software update on most phones as often the sampling rate is limited not by the sensing hardware, but by the operating system and sensor drivers to preserve power. Thus, a software update could change these driver values to provide a faster sampling rate to the operating system. The operating system can then perform some operation on the oversampled signal and then downsample the signal to the original sampling frequency. If the oversampled frequency is a multiple of the original, this can be trivially done by selecting x of y number of samples from the oversampled data. If a the oversampling frequency is a non-multiple, this could result in distortion of some kind being introduced into the digitized signal. This method provides the same signal sampling rate and bandwidth as current designs.

Digital anti-aliasing filters can employ oversampling to attenuate touchtone aliases while minimally altering other information applications may desire. The key is that the filters can remove any information above the original sampling rate's Nyquist frequency without effecting legitimate (i.e. not touchtone alias) data as seen in Fig~\ref{fig:defense-freq-aa}. Due to the non-linear nature of aliased frequencies, with oversampling the touchtone aliases may fall into range and can be attenuated without affecting benign information. This is not a panacea however, as aliases of sensitive information may still be in the original sampling range, but such a design can attenuate touchtone aliases without attenuating information hat applications may expect.

\subsubsection{Mitigations for targeted sensitive frequencies}
When there is a case of known sensitive signals with specific frequencies, such as in the case of mitigating touchtones, one can use frequency-specific mitigation designs such as notch filters and selective sampling frequencies in combination with anti-aliasing or other filtering techniques. A notch filter is a digital or analog filter design, similar to the high and low pass filters in Section~\ref{sec:background-signal-processing}, that attenuates information with frequencies between two cutoff frequencies. One could design multiple notch filters to attenuate targeted sensitive frequencies such as the eight touchtone frequencies.

\begin{figure}[!h]
    \centering
    \includegraphics[width=0.95\textwidth]{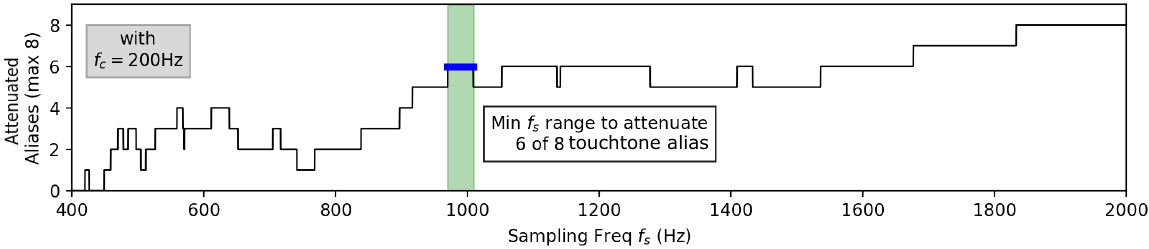}
    \caption{\textbf{Choosing a sampling rate to mitigate touchtone leakage.} A mitigation designer desiring to attenuate the most touchtone aliases using the lowest sampling rate $f_S$ when given a bandwidth $f_c$ to support can make use of the non-linear but predictable nature of aliased frequencies. As the oversampled rate increases, the number of aliased frequencies above $f_c$ will change. The designer can calculate this number of attenuated aliases then select an appropriate sampling rate to meet design constraints.}
    \label{fig:alias_count}
\end{figure}

Another approach is to use an anti-aliasing design while carefully selecting the sampling frequency to maximize the number of targeted sensitive frequencies above $f_N$ (Fig~\ref{fig:alias_count}). The basis for this lies in the non-linear relationship between signal frequency, sampling frequency, and the frequency alias as seen in Section~\ref{sec:background-signal-processing}. One can set the filter cutoff frequency $f_c$ to design for a desired bandwidth. Then, the designer could change the sampling rate until a desired number of aliases fall above $f_c$ so they can be eliminated. This could allow a mitigation designer to select lower sampling frequencies that result in greater protection from touchtone leakage.

\section{Experimental Method}\label{sec:setup}
To evaluate our mitigations we recorded touchtone samples on multiple phones with and without mitigations in place, then classified a test set of recordings to provide accuracy numbers.
Our machine learning classifier uses a variety of time and frequency features along with selective axis integration (as explained in Section~\ref{sec:attack-recovering}, to mimic a more advanced adversary. 
We use the same classifier as an evaluation metric to determine the accuracy of an adversary without any mitigations (a baseline) as well for both a baseline (without mitigations) and with mitigations in place.

\subsection{Data Collection}
\subsubsection{Hardware}
\begin{figure}[!h]
	\centering
        \includegraphics[width=.98\textwidth]{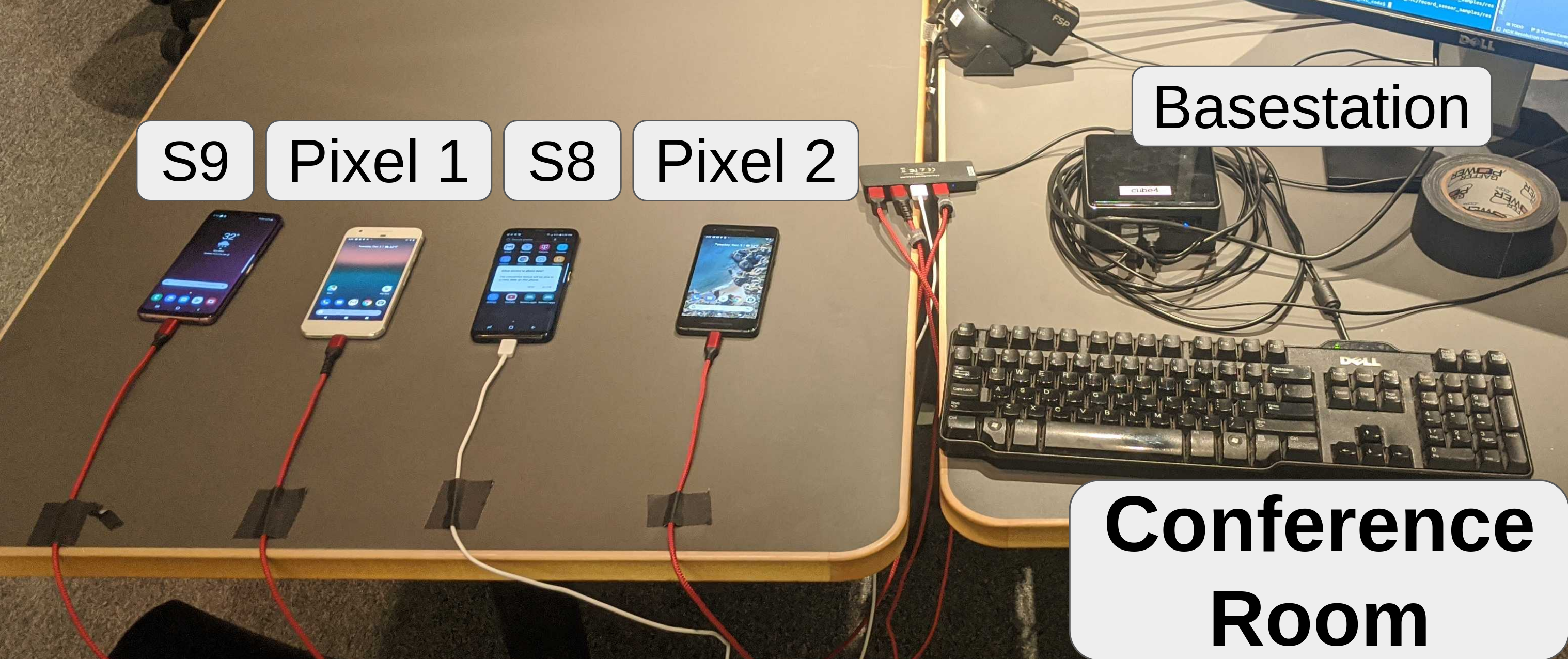}
        \caption{\textbf{Data collection setup in conference room.}}
        \label{fig:setup}
\end{figure}

We have three different hardware setups for motion sensor data collection. The first two setups collect data from the four Android phones listed in \ref{tab:phone_info} for baseline and software-only mitigation evaluation; these two setups differ only in physical locations: a quieter conference room versus a noisy server room. The conference room was next to a busy atrium with the door closed to mimic a conference call setting, while the server room was chosen to mimic a noisy environment measured at an average of 67~dB~SPL as measured by a General DSM403SD sound level meter\cite{dsm403sd}. Each setup used an Intel NUC running Ubuntu 18.04~\cite{nuc} as a base station, smart-phones (Table~\ref{tab:phone_info}), cables, and base station peripherals on a table~(Fig~\ref{fig:setup}). In this setup, the acoustic speaker was a phone's loudspeaker and the motion sensors (accelerometer and gyroscope) was the same phone's sensors. The base station used a python API for the Android Debug Bridge~\cite{adb} to upload a custom Android data collection program to each phone and for other communication or file transfer. 

The third hardware setup collects data at faster sampling rates for anti-aliasing software filters and for testing on-board sensor anti-aliasing filtering. Phone hardware can collect at rates faster than what is made available to applications in smartphones to limit power consumption.
Although current phones do not support it, we test it with external sensors to emulate a possible future mitigation. To that end, our setup uses a LSM9DS1 breakout boards, a very similar chip to the ones in three of the phones~(Table~\ref{tab:phone_info}), a Teensy 3.6 micro-controller, the same Intel NUC base station as in the previous setup, and an external speaker connected to the NUC to produce audio. The speaker was placed 10cm away from the LSM9DS1 breakout board. A python program was used to produce audio on the speaker and interface with a custom sensor collection program on the Teensy micro-controller.

\begin{table}[!h]
    \centering
    \caption{\textbf{Motion sensor information for phones used in experiments.} }
    \begin{tabular}{l | l | l | l |l}
    \hline
        \multirow{2}{2.0cm}{Manufacturer and Model} & \multirow{2}{1.2cm
        }{Release Date} & \multirow{2}{2cm}{Reported IMU Model} & \multicolumn{2}{c}{Sampling Rate (Hz)} \\ \cline{4-5}
        & & & Reported & Measured \\ \hline\hline
        Google Pixel 1 & Oct 2016 & BMI160 & 400.00 & 401.69 \\ \hline
        Google Pixel 2 & Oct 2017 & LSM6DSM & 400.00 & 409.96\\ \hline
        Samsung Galaxy S8 & Mar 2017 & LSM6DSL & 400.00 & 429.27 \\ \hline
        Samsung Galaxy S9 & Mar 2018 & LSM6DSL & 415.97 & 413.61 \\ \hline
    \end{tabular}\\
    Reported inertial measurement unit (IMU) model, which contains both an accelerometer and gyroscope, and sampling rates are found via the Android Debug Bridge tool. Note that the sampling rates are limited by the operating system, and not sensing hardware.
    \label{tab:phone_info}
\end{table}.

\subsubsection{Recording}

To reduce temporally correlated biases from data collection over a long period of time the \textit{python3} program running on the base station first determines a randomized order for all audio samples to record. The program then ensure proper setup of all devices for the experiment. It then has the speaker for the experiment play each touchtone audio clip in succession while recording motion sensor data. In the event with multiple devices connected to the base station, only one speaker and sensor were used simultaneously. Motion sensor data was collected at the fastest available sampling rate and saved and sent back to the base station to save the recording to disk.

For each individual setup we recorded the motion sensor data as each individual dial-tones was played for 0.5~s, with each tone being recording 250 times per setting for a total of 4000~recordings. The data set was divided into training and test sets at 80\% and 20\% respectively. It was ensured that touchtones were divided equally during the split (e.g. in the test set there was 50 samples of each of 16~touchtones).

\subsection{Touchtone classifier}
\begin{figure}[!h]
    \centering
    \includegraphics[width=0.95\textwidth]{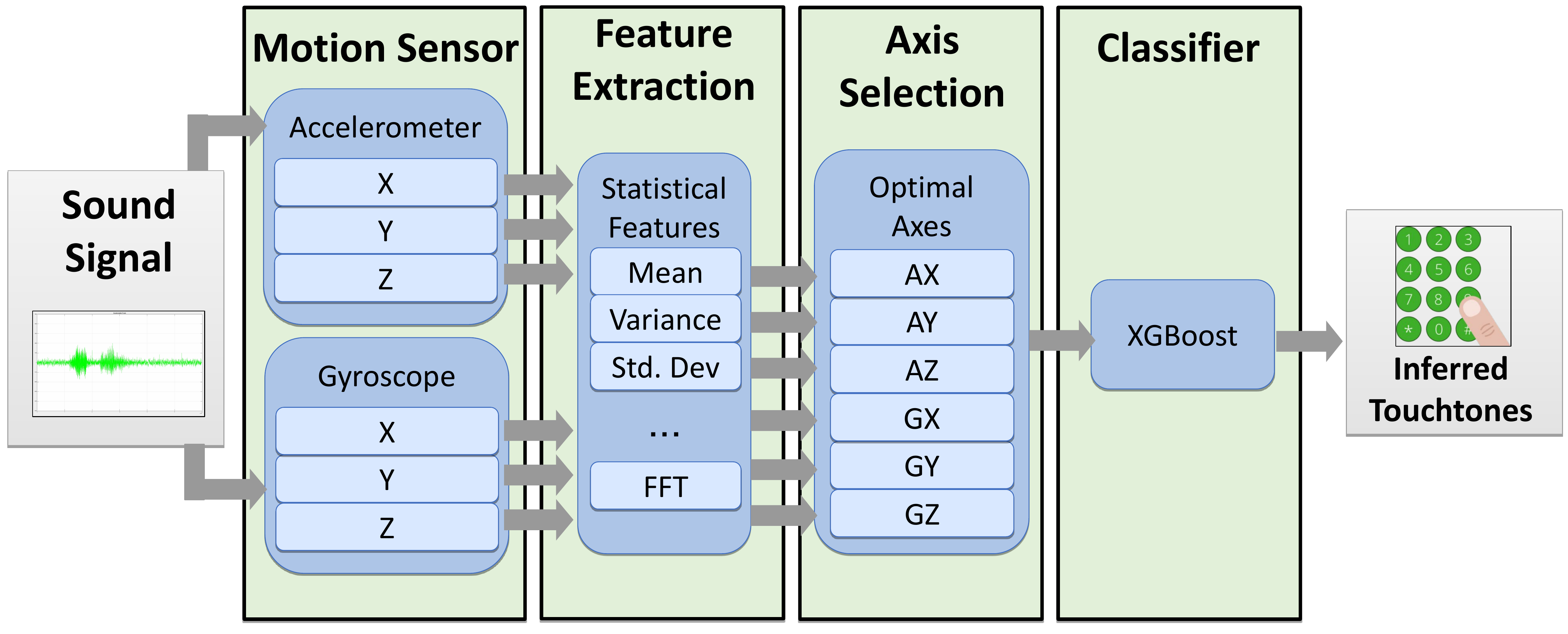}
    \caption{\textbf{Eavesdropping classifier.} Our system extract signals features and selectively combine useful motion sensor data from multiple sensors and axes to better classify touchtones.}
    \label{fig:attack_system}
\end{figure}

To serve as an evaluation metric we made a machine learning classifer~(Fig~\ref{fig:attack_system}) to mimic that of an advanced adversary.

\subsubsection{Selective integration of sensor data}
\label{sec:implementation-attack-fusion}
To emulate a more advanced adversary, we build a classifiers that
selectively integrates feature data from multiple sensors into a single attack model based on the intuition that each sensor axis can be a better or worse predictor for a given touchtone~(Section~\ref{sec:attack-recovering}). Previous work has demonstrated classifiers for acoustic leakage onto motion sensor~\cite{gyrophone,accelword,anand2019spearphone}, however to our knowledge no previous work has combined data from both sensors simultaneously or selectively integrated axes into a single model. This improvement works as each axis from each sensor carries some measure of unique information. Selectively combining these sources of unique information should yield best results.

Our method to selectively integrate axes is as follows. First the system imperially ranks the axes in order of best predictor by building a model for each individual axis and tests its accuracy on validation data. Then the system builds a model with the most accurate two axes, then top three, etc, until a model with all axes as been tested. Then the system selects the best performing model among the single-axis and multi-axis models to use in actual testing. Once the best combination of axes have been chosen, the axes will be selected in the Axis Selection step shown in Fig~\ref{fig:attack_system}.

\subsubsection{Features and Classifier Design}
\label{sec:implementation-attack-system}

We briefly detail the feature extraction and classifier of our touchtone classifier in this section. As a reminder, features are calculated per sensor axis, then features of only the optimal combination of axes are included in the model as described in Section~\ref{sec:implementation-attack-fusion}.

\smallskip\noindent\textbf{Time-alignment and Windowing:}
For feature extraction of a sample, our model first time-aligns signals from different sensors (i.e. sample 1 from one signal correlates with sample 1 of the others). Subsequently, it divides each time-series signal into a series of windows. Each window should correlate with windows of other signals (i.e. window 1 in one signal correlates with window 1 of another signal).

\smallskip\noindent\textbf{Extract Statistical Features:}
The system calculates a series of statistics per window per selected sensor axis, and concatenates these metrics to produce a single feature vector. The set of statistical measurements (Table~\ref{tab:time-freq-features}) are very similar to those used in previous work~\cite{anand2019spearphone}.
\begin{table}[!h]
    \centering
    \caption{\textbf{A list of features used in classification.}}
    \begin{tabular}{p{1.35cm} | p{1.4cm} | p{1.95cm} | p{2.95cm} |p{3.45cm}}
    \hline
        Mean & Median & Kurtosis & Absolute Area & \% Mean Crossings \\ \hline
        Minimum & Variance & Signal Power & Standard Deviation & Interquartile Range \\  \hline
        Range & Maximum & Variation & Spectral Entropy & Fast Fourier Transform \\  \hline
        Skew & \multicolumn{4}{c}{First, Second, Third Quantiles} \\ \hline
    \end{tabular}\\
     The signal would be split into windows where the above features were calculated.
    \label{tab:time-freq-features}
\end{table}

\smallskip\noindent\textbf{Zero-padding:}
Feature vectors with a different number of time windows, which may happen due to experimental error, must have the same number of features for the classifier to compare properly. The system zero-pads, or adds zeros or windows of zeros, each feature vector to ensure the same length.

\smallskip\noindent\textbf{XGBoost Classifier:}
Our system uses \textit{xgboost} to classify the extracted features from the selected axes. Xgboost is a common classifier that uses gradient boosting and has been shown to effective in several different applications~\cite{xgboost}.

\subsubsection{Implementation details and hyper-parameter tuning}
The system uses a \textit{python3} program to process the sensor recordings and subsequently train and/or test recognition models. We utilize numpy, scipy, and other standard \textit{python3} libraries to perform feature extraction as described previously. The system then uses \textit{python3} XGBoost implementation with support libraries from Scikit-learn to perform any training, validation, or testing of machine learning models. To select the optimal combination of axes as described previously, the system would first train separate models for individual axis. These axes would be then be ranked by individual accuracy performance and axes would be added in order of highest accuracy and evaluated. Last, for these eleven combinations (6~individual and 5~multi-axis) the system would chose the best performing axis combination and use that for its model.

To choose specific feature and model hyper-parameters, we performed a randomized grid search using data collected from a Pixel~2 phone in a conference room to imperially pick parameters. The randomized grid search did not test every possible combination of parameters in the interest of time, and thus it is possible more optimal parameters could be chosen. The possible parameters for features and classifiers are shown in Tables~\ref{tab:feature-settings}~and~\ref{tab:class-settings} respectively with selected parameters shown in bold. We tested these settings against a commonly used feature set for audio classification with Mel Frequency Cepstral Coefficients (MFCCs)~\cite{gyrophone} and another common classifier with Random Forest~\cite{anand2019spearphone} to provide a comparison against other commonly used selections. We took the highest accuracy result to select feature and classifier settings. These settings stayed the same through all testing.

\begin{table}[!h]
\caption{\textbf{Feature Settings. }}
    \centering
    \begin{tabular}{l | l| l}
    \hline
        Feature & Setting & Possible Choices \\ \hline \hline
        \multirow{2}{*}{\textbf{Statistic Features}} & Frame Size (\#vals) & 10, 20, \textbf{50}, 100 \\ \cline{2-3}
        & Frame Step (\#vals) & \textbf{5}, 10, 20\\ \hline \hline
        
        \multirow{2}{*}{MFCC} & Window Length (s) & 0.025, 0.05, 0.1, \textbf{0.2}, 0.3, 0.5 \\  \cline{2-3}
        & Window Step (s) & \textbf{0.01}, 0.05, 0.01 \\ \hline
        \end{tabular}
        \\
        The optimum feature settings used in the final model are in bold.  
    
    \label{tab:feature-settings}
\end{table}

\begin{table}[!h]
 \caption{\textbf{Classifier Settings.}}
    \centering
    \begin{tabular}{l | l | l}
    \hline
        Classifier & Setting & Possible Choices \\ \hline\hline

        \multirow{5}{*}{\textbf{xgboost}}

        & learning rate & 0.05, 0.10, 0.15, \textbf{0.20}, 0.25, 0.30 \\ \cline{2-3}
        & max depth &  3, 4, \textbf{5}, 6, 8, 10, 12, 15 \\ \cline{2-3}
        & min child weight &  1, \textbf{3}, 5, 7  \\ \cline{2-3}
        & gamma &  0.0, \textbf{0.1}, 0.2 , 0.3, 0.4  \\ \cline{2-3}
        & colsample bytree &  0.3, 0.4, \textbf{0.5} , 0.7 \\ \hline \hline
        
        \multirow{5}{*}{Random Forest}
        & bootstrap & True, \textbf{False}  \\\cline{2-3}
        & max depth & 10, 20, 30, 40, 50, 60, 70, \textbf{80}, 90, 100, None  \\\cline{2-3}
        & min samples leaf & 1, \textbf{2}, 4  \\ \cline{2-3}
        & min samples split & \textbf{2}, 5, 10  \\\cline{2-3}
        & \multirow{2}{*}{\textit{n}-estimators} & 200, 400, 600, 800, 1000, 1200, 1400 \\
        & & \textbf{1600}, 1800, 2000 \\
        \hline
        \end{tabular} 
        The optimum feature settings used in the final model are in bold.  
   
    \label{tab:class-settings}
\end{table}
\subsection{Signal Processing Mitigations}~\label{sec:setup-mitigations}
\subsubsection{Selection}
We selected a total of four signal processing mitigation designs to evaluate. Two designs, a software-only low pass filter and reduced sampling rates, were chosen as they are briefly mentioned in previous papers as possible mitigations for related work and may seem like natural mitigations for touchtone leakage. However, our analysis (Section~\ref{sec:mitigation-bad}) projects that both mitigations should have minimal effect on touchtone leakage without significantly reducing the available information to all applications. The low-pass filter used a Butterworth filter design with an order of 5.

The third and fourth mitigations, software and hardware digital anti-aliasing filters, were chosen to better support application functionality by not reducing bandwidth while still attenuating touchtone aliases. The tested software anti-aliasing filter design is essentially oversampling combined with filtering as shown in Fig~\ref{fig:defense-freq-aa}. Specifically, it uses the oversampled data with a Butterworth low-pass filter and we test various filter orders. The Butterworth filter provides a good balance with only slight signal distortion and a sharper cutoff. The slight signal distortion means it should minimally affect applications relying on sensor data while the sharper cutoff means it should theoretically attenuate aliased signals further. Furthermore, this filter can be implemented as a software update to any phone, but will require some computational burden and cause some signal delay, which may be unacceptable for some applications.

The hardware digital anti-aliasing filter refers to the on-board anti-aliasing filter on the LSD9DS1 breakout board, which should also be included on the LSM6DS(L/M) sensors on three of the tested phones. This filter should work similarly in theory to software anti-aliasing filter as they are both digital anti aliasing filters, but the exact filter details are unfortunately black-box. The hardware filter benefits over the software version in that it requires no computational burden and should require less signal delay, but has the drawback that is less configurable and may not be available on some devices. To note, it \textit{can} be implemented as a software update should the hardware be available by changing values in the sensor driver.

\subsubsection{Implementation Details}
Implementation details for our four tested mitigations include:
\begin{enumerate}
    \item \textbf{Reduced sampling rate.} The reduced sampling rate mitigation uses the original motion sensor data from the conference room hardware setup but takes 1 sample of every $n$ samples to emulate the effect of reducing sensor sample rate by $n$. We vary $n$ to test sampling rates from 400~Hz to as low as 50~Hz, with Nyquist frequency and bandwidth equal to half the sampling rate.
    
    \item \textbf{Software low-pass filtering.} The low-pass filter uses the original motion sensor data from the conference room hardware setup with an unaltered sample rate but applied the Python scipy Butterworth filter with an order of 5 for low-pass filtering. The signal cut-off frequency was varied from 200, 150, 100 to 50 Hz. This cut-off frequency effectively becomes the bandwidth of unattenuated information in the signal. 

    \item \textbf{Software (digital) anti-aliasing filtering.} The software anti-aliasing filter uses the oversampled data from the sensor breakout board setup and then applies a scipy Butterworth filter (the same as from the original low-pass filter) with a cutoff frequency equal to the eventual desired bandwidth. We vary this desired bandwidth to use as comparison against other mitigations. The filtered signal is then downsampled (similarly to the reduced sampling rate mitigation) to the desired bandwidth. We also vary filter order in this evaluation.
    
    \item \textbf{Hardware (analog) anti-aliasing filtering.} The hardware anti-aliasing filter collects data from the sensor breakout board, changing the sensor's on-board filtering settings. There are four bandwidth configurations. The  data from the four configurations is then processed by the classifier.
    
\end{enumerate}

\section{Evaluation Results and Analysis}

In this section we report the attack and mitigation results with the setups described in Section~\ref{sec:setup}. We analyze and summarize the findings of our assessment of the eavesdropping attack and different mitigations.

Software low-pass filtering and reducing the sensor sampling rate can only moderately mitigate the attack while significantly hindering data bandwidth (and thereby application functionality). Software and hardware digital anti-aliasing filters cannot eliminate touchtone eavesdropping, but are able to more significantly mitigate the threat while also preserving more data bandwidth. 

\subsection{Baseline evaluation metrics: attack effectiveness}
\begin{figure*}[!h]
	\centering
    \includegraphics[width=.95\textwidth]{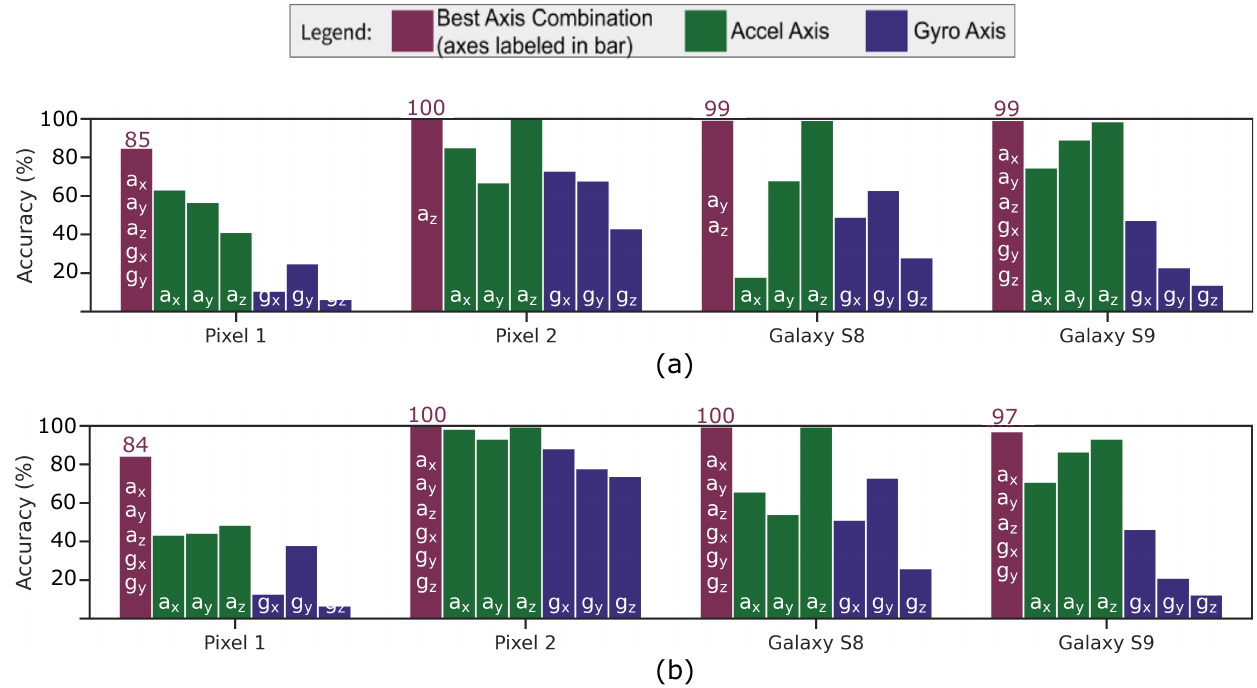}
    \label{fig:eval-attack-dtmf-server}
	\caption{\textbf{Baseline results for the touchtone eavesdropper.} (a) Conference room and (b) Server room hardware setups. For each phone, we show the accuracy a classification model trained on individual axes alone, then show the accuracy for the model trained on the optimal combination of axes.}
	\label{fig:eval-attack}
\end{figure*}

We find that the unmitigated touchtone classifier achieves accuracy exceeding 99\% for three of the four phones as shown in Fig~\ref{fig:eval-attack}, demonstrating that malicious applications can effectively recover user input.

\subsubsection{Differences between phone models} \label{sec:eval_results_phones}
One of the phones, the Pixel~1, performs poorest in nearly every test despite similar sampling rates as the other phones. The highest touchtone inference accuracy for Pixel 1 does not exceed 85\% while other phones can all achieve over 99\%. The most obvious explanation would be that it has different inertial measurement unit produced by a different manufacturer than all other phones (Table~\ref{tab:phone_info}).

This result demonstrates that factors other than sampling rates can vary recognition rates. These factors could include: signal propagation path that attenuates the acoustic signal, less sensitive sensors, different frequency responses, or different sensor configurations. This result also suggests that some motion sensors may be more resistant to touchtone leakage than others. An examination on which motion sensors are less susceptible could provide insight into future hardware-based mitigations.

\subsubsection{Accelerometer vs. gyroscope axis accuracies}
Classification based on data from an accelerometer axis achieved higher average accuracy gyroscope axis data.
While the exact reasons remain unclear, we provide a possible assumption. Accelerometers measure linear acceleration while gyroscopes measure angular acceleration. The phone's speakers produce audio through vibration, and then vibration travels through the phone body to affect both the accelerometers and gyroscopes. Vibration acts as linear acceleration in this case, which the accelerometer is designed to measure. While the gyroscope is not designed to measure linear acceleration, its sensing mass(es) still vibrate and these vibrations are quantized. Thus, the intent of each sensor changes the effectiveness for this particular scenario.

\subsubsection{Selective integration of sensor axes.}
Selective integration of axis data only achieved significantly higher results for one phone model, the Google~Pixel~1, but it did improve accuracy versus a single axis for all but one case. This case was the test for the Google~Pixel~2 in the conference room, and it could not improve accuracy as accuracy was already 100\%. For all phones but the Pixel~1, the improvement was limited because the results were already near 100\% accuracy. However, for the Pixel~1 the selective-axis integration improved as much as 40\% over single-axis accuracies. This indicates that in cases with noisier data, the selective axis integration could help a classifier model utilize the various touchtone information in each axis to achieve higher accuracies.

\subsection{Mitigation strategy evaluations}

\subsubsection{Reduced Sampling Rates} \label{sec:results-reduced}
\begin{figure*}[!h]
	\centering
    \includegraphics[width=.9\textwidth]{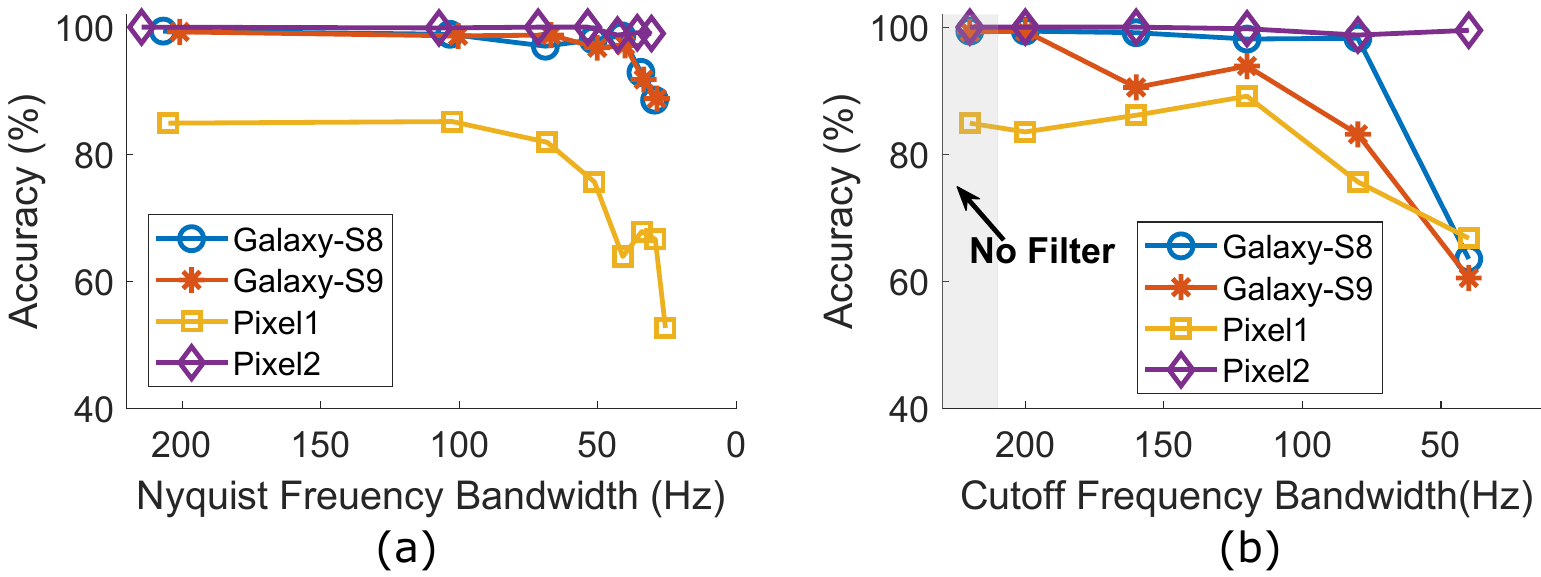}
	\caption{\textbf{Functionality-unaware mitigation results.} (a) Results for a downsampling mitigation with listed bandwidth equivalent to half the used sampling rate. (b) Results for the low-pass filter mitigation with listed bandwidth equivalent to the cutoff frequency used. Both mitigations do not greatly reduce touchtone eavesdropping accuracy until bandwidth is under 50~Hz, which could hinder functionality for benign applications.}
	\label{fig:eval-ineffecitve}
\end{figure*}

The results for the reduced sampling rates mitigation support the theoretical analysis in Section~\ref{sec:mitigation-reduced} to show that this approach does not greatly affect touchtone eavesdropping until sampling rates are reduced significantly~(Fig~\ref{fig:eval-ineffecitve}a). Once again, this is because touchtone aliases will remain in the digitized signal no matter the sampling rate, and this mitigation affects eavesdropping accuracy by reducing the total information available (affected functionality of benign applications). More concretely in our results the reduced sampling rate does not seem to have much of an effect until the sampling frequency is roughly 100~Hz, and thus bandwidth reaches around 50~Hz.

\subsubsection{Software Low-pass Filter} \label{sec:results-lpf}
An evaluation of software low-pass filter shows that, as expected~(Section~\ref{sec:mitigation-lpf}), they may also do not seem to affect accuracy significantly until lower bandwidths are researched. As Fig~\ref{fig:eval-ineffecitve}b demonstrates, in our tests the touchtone accuracy results were only minimally affected until a very low 40~Hz cutoff frequency was reached. In fact in some cases, such as with the Pixel~1's accuracy results, the average accuracy actually improved when using cutoffs of 160~Hz and 120~Hz, which could indicate a large amount of noise in the 160~Hz to 200~Hz range for that particular phone. However, for the Pixel~2 the accuracy for touchtones remained essentially unaffected even at a 40~Hz cutoff frequency.

\subsubsection{Software and Hardware Anti-aliasing Filter} \label{sec:results-aa}

\begin{figure*}[!h]
	\centering
    \includegraphics[width=.8\textwidth]{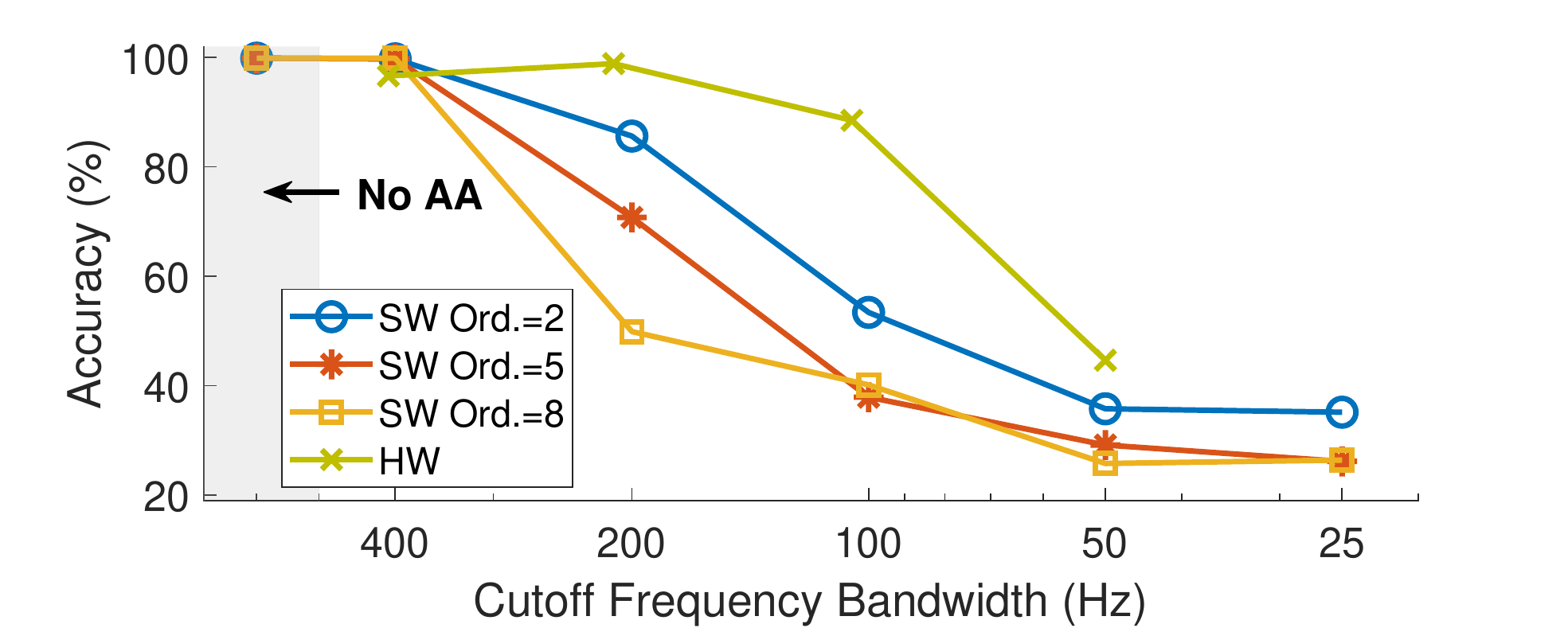}
	\caption{\textbf{Anti-aliasing results.} While not completely eliminating the attack, the software anti-aliasing filter is able to significantly reduce the accuracy of the touchtone eavesdropper and speech snooper attacks. Additionally, with better hardware implementations (higher sampling rates), this mitigation could be even more effective while preserving bandwidth.}
	\label{fig:eval-def-aa}
\end{figure*}

Our experiments show that anti-aliasing filters are more effective than either pure low-pass filters or sampling rate reduction. For example, at with a cutoff frequency/bandwidth of 100 Hz, the order 5 and 8 software anti-aliasing filters reduce the touchtone eavesdropping accuracy from over 99\% to below 40\% (see Fig~\ref{fig:eval-def-aa}), while neither the pure low-pass filter nor the reducing sampling rate mitigation could reduce the accuracy to below 80\%. 

Both the software and hardware anti-aliasing filter work to some degree, but neither are perfect mitigations. Most manufacturers may expect a built-in, hardware based ``anti-aliasing filter" to fully filter aliases without looking into the details. While somewhat effective, it was actually the software based solution we found to work best. We do not currently know the exact filter parameters of the hardware filter due to the black box nature of the design. However, the hardware implementation would likely increase in effectiveness by increasing filter order, as seen in our software implementation. The 8th order, 200~Hz bandwidth software anti-aliasing filter is likely the best solution from our experiments as it preserves the 200~Hz bandwidth used by most of our phones while still having a significant effect on the attacks.
\section{Discussion}
\subsection{Hardware solutions}
In this paper we do not evaluate sensor hardware changes for mitigations as they would require mitigations as they cannot be implemented as a software update.
However, some mitigations embedded into circuitry could serve as long-term solutions. Analog filtering mitigations schemes should work well against touchtone leakage as they can directly attenuate the original touchtone frequencies before sampling, and therefor before aliasing. Additionally, randomized sampling could be used to mitigate some of the aliasing effects.

\subsection{Application to other acoustic leakage}
The analysis of the paper focuses on touchtones, but it is largely applicable to other forms of acoustic leakage. We focus on touchtone leakage as it provides a high-impact, yet simple signal for attack and mitigation analysis when compared to other potential targets such as speech. Touchtones, while simple, are still difficult to mitigate and possibly more difficult than speech as attackers likely need less information to classify touchtones over complex signals such as speech. Thus, mitigation strategies that work for touchtones should largely work for speech.

\section{Conclusion}
This paper examines how to mitigate touchtone leakage-enabled acoustic eavesdropping with smartphone motion sensors.
A smartphone's motion sensor data has information of which touchtones are pressed by a victim when conducting activities such as dialing a phone number, navigating an automated telephony service, or activating a credit card number.
An adversary understanding the relevant physics and signal processing concepts such as signal aliasing and  different frequency responses can use motion sensor data to infer the unique touchtones and thus the user's input.
Some of the more obvious mitigations, such as software low-pass filters or reduced sampling rates, may actually have very little effect on mitigating touchtone leakage.
We instead propose software and hardware digital anti-aliasing filtering designs which achieve moderate success and can be implemented as a software update.

\nolinenumbers

%
%
%
\bibliography{bibliography.bib}

\end{document}